\documentclass{amsart}

\RequirePackage{fix-cm}
\usepackage{graphicx}
\usepackage{amssymb}
\usepackage{bm}
\usepackage{tikz}
\usepackage{endfloat}
\usepackage{setspace}
\newcommand{\eref}[1]{(\ref{#1})}
\newcommand{\matr}[1]{\mathbf{#1}}
\newcommand{\vect}[1]{\mathbf{#1}}

\newtheorem{theorem}{Theorem}[section]

\theoremstyle{definition}
\newtheorem{definition}[theorem]{Definition}

\theoremstyle{remark}

\begin{document}

\title[A conceptual framework for discrete IP in geophysics]{A conceptual framework for discrete inverse problems in geophysics
}


\author[M Giudici et al.]{Mauro Giudici}
\thanks{Discussions with several students and researchers worldwide seeded a lot of the concepts discussed in this paper and they are gratefully acknowledged.}

\author[]{Fulvia Baratelli}

\author[]{Laura Cattaneo}

\author[]{Alessandro Comunian}

\author[]{Giovanna De Filippis}

\author[]{Cinzia Durante}

\author[]{Francesca Giacobbo}

\author[]{Silvia Inzoli}

\author[]{Mauro Mele}

\author[]{Chiara Vassena}

\address{M.~Giudici, L.~Cattaneo, A.~Comunian,  C.~Durante, S.~Inzoli, M.~Mele, C.~Vassena, Universit\`a degli Studi di Milano, Dipartimento di Scienze della Terra ``A.Desio'', via Cicognara 7, I-20129 Milano, Italy}
\curraddr{}
\email{mauro.giudici@unimi.it}
\address{F.~Baratelli, MINES ParisTech - Centre de G\'eosciences, 35 rue Saint Honor\'e, 77305 Fontainebleau, France}
\address{G.~De Filippis, Istituto di Scienze della Vita, Scuola Superiore Sant'Anna, Via S.~Cecilia 3, I-56127 Pisa, Italy}\curraddr{AECOM URS ITALIA S.p.A.}
\address{F.~Giacobbo, Politecnico di Milano, Dipartimento di Energia, Sezione Ingegneria Nucleare, Piazza L. da Vinci 32, 20133 Milano, Italy}

\date{first version January 23 2019; last revision \today}

\begin{abstract}
In geophysics, inverse modelling can be applied to a wide range of goals, including, for instance, mapping the distribution of rock physical parameters in applied geophysics and calibrating models to forecast the behaviour of natural systems in hydrology, meteorology and climatology. A common, thorough conceptual framework to define inverse problems and to discuss their basic properties in a complete way is still lacking. The main goal of this paper is to propose a step forward toward such a framework, focussing on the discrete inverse problems, that are used in practical applications. The relevance of information and measurements (real world data) for the definition of the calibration target and of the objective function is discussed, in particular with reference to the Bayesian approach. Identifiability of model parameters, posedness (uniqueness and stability) and  conditioning of the inverse problems are formally defined. The proposed framework is so general as to permit rigorous definitions and treatment of sensitivity analysis, adjoint-state approach, multi-objective optimization.
\end{abstract}


\maketitle

\section{Introduction}\label{intro}
Mathematical models of geophysical processes and phenomena represent useful tools for different goals. They can be applied to interpret the results of field or laboratory measurements, to set up monitoring networks and experimental devices and procedures, to forecast the behaviour of geophysical systems under different stresses and conditions of exploitation of natural resources, to assess the environmental impact of buildings and infrastructures, to perform risk analysis related to natural hazards, to design measures for remediation of contaminated sites, etc. Most geophysical processes can be mathematically represented by means of partial differential equations, but the inventory of equations used in different situations is very wide and a great number of solution methods is applied.

When mathematical models are applied to practical problems, an accurate estimate of model parameters is fundamental and measurements are essential to calibrate numerical models by solving inverse problems. Therefore, the properties of inverse problems strongly depend on data collection and processing.

While the goal of inverse problems is to determine some of the model parameters, the objectives of inverse modelling are very widespread and the following examples can be recalled:
\begin{enumerate}
	\item mapping the distribution of physical parameters in the subsurface is required for geological studies and to solve practical problems in the fields of: civil, environmental, and geological engineering; hydrogeology; exploration of mineral resources and hydrocarbon reservoirs; geoarcheology and cultural heritage studies; etc.; 
	\item finding the optimal parameters is necessary to reliably model the evolution of natural systems in response to changes of the stresses, whose origin can be artificial (e.g., exploitation of natural resources for human needs) or natural (e.g., climate changes);
	\item fitting simple models to laboratory data taken on samples is useful to characterize the behaviour of the materials of geophysical interest (rocks, water, ice, air);
	\item and so on.
\end{enumerate}

On the basis of the previous remarks, the objective of this paper is
to propose a common, formal and conceptual framework, which permits to
define inverse problems (IPs) and discuss their properties in the
different fields of geophysics. The proposed framework allows to handle the great variety of relevant mathematical models and, in
particular, permits to discuss the role played by the following
factors: expected use of the model; data collection and processing;
methods of discretization of partial differential equations; methods
of solution of discrete equations; deterministic or stochastic
approaches. In this paper, the attention is restricted to the case of
discrete IPs, because practical problems always require numerical
computations.

Several papers and textbooks \cite{Aster2013,Menke2012,Parker1994,Tarantola2004,Zhdanov2015} introduce general definitions of IPs. 
Nevertheless, there is room and need for a more comprehensive and flexible conceptual framework, which should allow to cast the definition and properties of inverse modelling in a more precise way.

It is expected that the proposed conceptual framework provides a better insight in the role that data have on the model calibration, including their use to estimate the target values of the physical quantities which are compared with model predictions. In particular, the classical definition given in textbooks assumes that the solution to the forward problem (FP) is directly compared with measurements. However, the strict outcome of the FP, which is often the state of the system,
in many situation cannot be directly compared with the available measurements. Instead, it is often used to compute other model predictions, which can then be compared with measured data.
On the other hand, the inversion target could be obtained by processing the measurements. Therefore, one of the goals of the proposed framework is to highlight the different role of FP output, model predictions, measured quantities and target values. As a consequence, this conceptual framework could help to focus the role of data on the development and the application of a model.

Moreover, the proposed framework is designed to clearly distinguish definitions and different properties that could be intrinsically related either to the FP or to the IP from properties that might depend on the algorithm used to solve the IP. While some definitions given in this paper simply replicate those found in textbooks, here the discussion of fundamental issues related to the posedness of the IP (uniqueness, stability, conditioning) is much more developed.

Two paradigmatic examples will be adopted for these purposes.

\section{A conceptual framework for mathematical modelling and inverse problems in geophysics}

\subsection{Paradigmatic examples}

Two paradigmatic examples are considered in this paper, in order to facilitate the description of the proposed conceptual framework and to provide instances of its application. The first example is a simple scheme of cross-hole seismic tomography and is a prototype of a linear model, which is defined as a model for which the state of the system linearly depends on the model parameters to be calibrated. This example is useful to discuss some of the properties of the IP for exploration geophysics. The second example is related to the study of diffusive processes under stationary conditions, which is the paradigm of the discrete counterpart of the IP of estimating the leading coefficient of an elliptic partial differential equation. This problem finds application in th estudy of several geophysical processes, for instance groundwater circulation, heat transfer and solute transport.

\subsubsection{Example 1: cross-hole seismic tomography}

\begin{figure}[htbp]
	\centering
		\includegraphics[width=10cm]{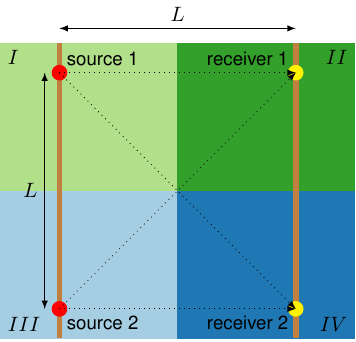}
	\caption{Geometrical sketch for the paradigmatic example 1. Brown lines: boreholes. Dots: sources (red) and receivers (yellow). The four coloured boxes, labelled with Roman numbers, represent volumes characterized by different values of seismic-waves propagation velocity. Dotted lines: straight path of the seismic waves from sources to receivers (refraction is neglected).}
	\label{fig:Figure01}
\end{figure}

Suppose that two boreholes are drilled at a distance $L$ and that one of them is equipped with two sources of seismic waves, whereas the other one is equipped with two receivers of seismic waves (Fig.~\ref{fig:Figure01}). For the sake of simplicity, assume that the depths at which the two sources are located in the first borehole are the same at which the two receivers are located in the second borehole. In particular, assume that the distance of sources inside the first borehole is equal to the distance between the receivers in the second borehole and equal to $L$. Under this configuration, four measurements can be taken, i.e. the traveltimes $t_{m,n}$ needed by seismic waves to start from one of the sources, $m$, and reach one of the receivers, $n$.

Thanks to the simple geometry considered in this example (Fig.~\ref{fig:Figure01}) and by neglecting the refraction related to Snell's law, it is possible to set up a very simple model, which considers straight paths of the seismic waves from a source to a receiver, travelling through four blocks of the subsurface characterized by different values of propagation velocity of seismic waves ($V_I$, $V_{II}$, $V_{III}$, $V_{IV}$):
\begin{equation}\label{eq:CompleteModel}
\begin{array}{rcl}
	t_{1,1}^\mathrm{(mod)} &=& L(2V_{I})^{-1}+L(2V_{II})^{-1},\\
	t_{1,2}^\mathrm{(mod)} &=& L\sqrt{2}(2V_{I})^{-1}+L\sqrt{2}(2V_{IV})^{-1},\\
	t_{2,1}^\mathrm{(mod)} &=& L\sqrt{2}(2V_{II})^{-1}+L\sqrt{2}(2V_{III})^{-1},\\
	t_{2,2}^\mathrm{(mod)} &=& L(2V_{III})^{-1}+L(2V_{IV})^{-1}.
\end{array}
\end{equation}
For this example, the FP aims at determining the travel times $t_{m,n}$, given the values of the propagation velocities, whereas the IP aims at finding the propagation velocity of seismic waves in the four blocks.

\subsubsection{Example 2: stationary diffusion}

Diffusive processes are modelled with partial differential equations which are based on physical conservation principles (e.g., mass, energy, linear momentum) and on phenomenological laws (Fick's, Fourier's, Darcy's, etc.) and which are complemented with boundary and initial conditions (BICs). Discrete models are designed by discretizing these equations and the BICs with a large number of techniques (e.g., finite differences, finite elements, spectral methods); the final result is a set of algebraic, possibly non-linear, equations.

\begin{figure}[htbp]
	\centering
		\includegraphics[width=10cm]{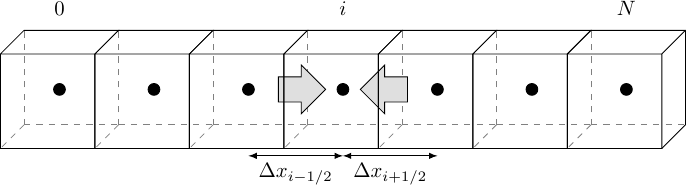}
	\caption{Discretization grid for the paradigmatic example 2. Nodes are denoted as black dots; arrows show the incoming fluxes through the borders of the cell $i$, appearing in the left-hand-side of \eref{eq:FiniteDifference}.}
	\label{fig:Figure02}
\end{figure}

In the simplest case of a 1D, purely diffusive (i.e., convective terms are neglected), stationary process, a conservative finite-difference approximation can be synthetically written as
\begin{equation}\label{eq:FiniteDifference}
a_{i-1/2}\frac{u_{i-1}-u_{i}}{\Delta x_{i-1/2}} + a_{i+1/2}\frac{u_{i+1}-u_{i}}{\Delta x_{i+1/2}} = \varphi_i,\ i=1,\ldots,N-1,
\end{equation}
where: $i\in\{0,\ldots,N\}$ is the index used to identify a node, which is the centre of one of the $N+1$ non-overlapping cells that cover the whole domain (see the geometry in Fig.~\ref{fig:Figure02}); $a_{i-1/2}$ represent the phenomenological coefficients (e.g., the product of internode or interblock conductivity times the surface separating two adjacent blocks); $u_i$ is the potential at node $i$ (e.g., water head, solute concentration, or temperature); $\Delta x_{i-1/2}$ is the spacing between adjacent nodes; and $\varphi_i$ is the sum of the source terms (expressed as a flow rate of the considered quantity per unit surface) in the cell $i$.

Notice that each of the terms in the left hand side of \eref{eq:FiniteDifference} represents the specific flux per unit surface of the considered quantity (e.g., mass, energy) entering the cell $i$ through the surface separating two adjacent cells.

Dirichlet boundary conditions are easily introduced, by setting $u_0$ or $u_N$ equal to the prescribed value. Neumann boundary conditions can be introduced by substituting $a_{1/2}\left(u_{0}-u_{1}\right)/\Delta x_{1/2}$ with the prescribed value $q_{1/2}$, or analogously for the term $a_{N-1/2}\left(u_{N}-u_{N-1}\right)/\Delta x_{N-1/2}$ with $q_{N-1/2}$. Notice that Neumann boundary conditions permit to write \eref{eq:FiniteDifference} also for $i=0$ or $i=N$.

For this example, the FP aims at solving \eref{eq:FiniteDifference} with respect to $u_i$, if $a_{i-1/2}$ and $\varphi_i$ are known, whereas the IP aims at identifying the best values of the phenomenological parameters $a_{i-1/2}$, $i=1,\ldots,N-1$.

\subsection{Basic definitions}\label{sec:BasicDefinitions}
Any discrete mathematical model can be represented by a set of equations that describe the state of the physical system under study as a function of model parameters. The model parameters are included in an array $\vect{p}$, whereas an array $\vect{s}$ includes the quantities that describe the state of the system.

Notice that the term ``array'' is used to designate a collection of physical quantities and parameters that will be considered either as a column vector to which one can apply the methods of linear algebra or as a finite set of elements.

In the most general form, a discrete model can be written as the following system of equations
\begin{equation}\label{eq:DiscreteModel}
	\vect{f}(\vect{p},\vect{s}) = 0,
\end{equation}
where the functions $\vect{f}$ may assume different forms for different problems, as shown in Secs.~\ref{subsub-Ex1} and \ref{subsub-Ex2} for the two paradigmatic examples.

Roughly speaking, the FP aims at solving \eref{eq:DiscreteModel} with respect to $\vect{s}$, given the model parameters $\vect{p}$, whereas the IP aims at identifying the values of some of the model parameters. If the numbers of model parameters and of state parameters are, respectively, $N_\mathrm{(p)}$ and $N_\mathrm{(s)}$, then $\vect{p}\in\mathcal{P}\subseteq\mathbb{R}^{N_\mathrm{(p)}}$ and $\vect{s}\in\mathcal{S}\subseteq\mathbb{R}^{N_\mathrm{(s)}}$, where the subspaces $\mathcal{P}$ and $\mathcal{S}$ could take into account some physical constraints on the model parameters and the state variables. As a simple example, conductivities should be non negative.

From a more formal point of view, one can give the following
\begin{definition}[Forward problem -- FP] Let $\vect{p}\in\mathcal{P}\subseteq\mathbb{R}^{N_\mathrm{(p)}}$ be known; then the FP is defined as finding $\vect{s}\in\mathcal{S}\subseteq\mathbb{R}^{N_\mathrm{(s)}}$ such that \eref{eq:DiscreteModel} is satisfied.
\end{definition}

If a unique solution of the FP can be found, it can be expressed in explicit, possibly non-linear form as
\begin{equation}\label{eq:ExplicitModel}
	\vect{s}=\vect{g}(\vect{p}).
\end{equation}
For the sake of simplicity, in this paper it is assumed that the FP is well-posed, i.e., a solution exists, is unique and depends with continuity on $\vect{p}$. However, notice that this is not always the case \cite{Valota2002}.

The array $\vect{p}$ includes any model parameter, comprising those which describe the geometry of the discretization grid (e.g., the spacing of the grid or the time step for time-evolving processes). Therefore, some of these parameters are fixed before the application of the model; their values will depend on the available data, which are the elements of the array $\vect{d}$. Then the fixed parameters can be grouped in a ``sub-array'' $\vect{p}^\mathrm{(fix)}$, which depends on the data: $\vect{p}^\mathrm{(fix)}(\vect{d})$.

The array $\vect{p}^\mathrm{(cal)}\in\mathbb{R}^{N_\mathrm{(c)}}$ is used to denote the model parameters, whose values are obtained from the solution of an IP. The number of elements of $\vect{p}^\mathrm{(cal)}$ is $N_\mathrm{(c)}$. Therefore
\begin{equation}
\vect{p}=\left({\vect{p}^\mathrm{(fix)}}^t, {\vect{p}^\mathrm{(cal)}}^t\right)^t.
\end{equation}

A model is said to be linear if \eref{eq:ExplicitModel} can be expressed as follows
\begin{equation}\label{eq:LinearModel}
	\vect{s}=\matr{G}\vect{p}^\mathrm{(cal)},
\end{equation}
where $\matr{G}\left(\vect{p}^\mathrm{(fix)}\right)$ is an $N_\mathrm{(c)} \times N_\mathrm{(s)}$ matrix, which is independent of $\vect{p}^\mathrm{(cal)}$.

The model outcome, i.e., the state of the system, can be used to forecast other quantities that might depend on the model parameters and possibly on some of the data. For instance, with reference to the paradigmatic example 2, flow rates could be computed by using the state of the system and the internode conductances. Therefore, the model forecast is expressed as an array $\vect{y}$, which is function of $\vect{s}$, $\vect{p}$ and $\vect{d}$: $\vect{y}\left(\vect{s},\vect{p},\vect{d}\right)$.

Roughly speaking, the IP consists in the determination of the optimal values of $\vect{p}^\mathrm{(cal)}$, that are the values that reduce the misfit between model forecasts and target values. In the simplest case, if measurements of the state of the system were available, this would imply that some elements of $\vect{s}$ should be directly compared with the corresponding elements of $\vect{d}$. Unfortunately, this is an oversimplification of what is done in practice. Therefore, it is necessary to introduce the calibration target. The latter is the array $\vect{t}$ that collects the values which should be attained by the model forecast, if the model were physically ``correct'' and the model parameters were ``optimal''. The array $\vect{t}$ may depend on $\vect{d}$ and $\vect{p}^\mathrm{(fix)}$, but should be independent of $\vect{p}^\mathrm{(cal)}$: $\vect{t}=\vect{t}\left(\vect{d},\vect{p}^\mathrm{(fix)}\right)$. Let $N_\mathrm{(t)}$ denote the number of elements of $\vect{y}$ and $\vect{t}$.

\subsubsection{Example 1}\label{subsub-Ex1}

For example 1, the data include the measured traveltimes ($t_{m,n}$), but also the positions of the sources and the receivers. Here, the unknown parameters are the seismic-wave propagation velocities of the four blocks in which the subsurface is subdivided. Equation \eref{eq:CompleteModel} can be rewritten in the form of a linear model \eref{eq:LinearModel}, if $\vect{s}$ includes $t_{m,n}^\mathrm{(mod)}$ from \eref{eq:CompleteModel}, 
\begin{equation}\label{eq:mcalEx1}
\vect{p}^\mathrm{(cal)}=\left(V_I^{-1},V_{II}^{-1},V_{III}^{-1},V_{IV}^{-1}\right)^t,
\end{equation}
and the matrix $\matr{G}$ is given by
\begin{equation}\label{eq:GEx1}
\matr{G}=\frac{L}{2}\left( 
\begin{array}{cccc}
	1        & 1        & 0        & 0        \\
	\sqrt{2} & 0        & 0        & \sqrt{2} \\
	0        & \sqrt{2} & \sqrt{2} & 0        \\
	0        & 0        & 1        & 1
\end{array}
\right).
\end{equation}

In other words, the physical parameters to be identified are the inverse of the propagation velocity of seismic-waves (called ``slowness'' in the jargon of seismic prospecting). Notice also that the elements of $\matr{G}$ depend on the positions of the sources and the receivers; these data, for the simple geometry that is considered in this paper, are ``concentrated'' in the parameter $L$. Therefore, the positions of the sources and the receivers are included in the $\vect{p}^\mathrm{(fix)}$ array and are estimated from a subset of the data array $\vect{d}$. So, $\matr{G}$ is a function of $\vect{p}^\mathrm{(fix)}$.

For this example, the target array simply includes the measured traveltimes, so that it is a subset of the array $\vect{d}$:
\begin{equation}\label{eq:targetEx1}
\vect{t}=\left( t_{1,1}, t_{1,2}, t_{2,1}, t_{2,2} \right)^t.
\end{equation}
On the other hand, the model prediction array $\vect{y}$ simply coincides with $\vect{s}$.

\subsubsection{Example 2}\label{subsub-Ex2}

The state array $\vect{s}$ corresponds to the potential at the internal nodes, that are the nodes for which the balance equation \eref{eq:FiniteDifference} can be written.

If \eref{eq:FiniteDifference} is written for every interior node, and if it is completed with boundary conditions, then a system of linear equations is obtained, which can be written as follows:
\begin{equation}\label{eq:prototype}
	\matr{A}\left( \vect{p} \right) \vect{s} = \vect{b}\left( \vect{p} \right),
\end{equation}
where $\matr{A}$ is a square matrix and $\vect{b}$ is used to model both the source terms and the boundary conditions. In this case, \eref{eq:DiscreteModel} becomes
\begin{equation}\label{eq:prototype_f}
	\vect{f}\left( \vect{p}, \vect{s} \right) = \matr{A}\left( \vect{p} \right) \vect{s} - \vect{b}\left( \vect{p} \right) = 0.
\end{equation}

Equation \eref{eq:prototype} is a prototype also for the linear system of equations arising from the discretization of the partial differential equations for 2D or 3D problems. It is a prototype also for transient conditions, for which the array $\vect{s}$ is usually split in the sub-arrays corresponding to different time steps. Also, \eref{eq:prototype} is a prototype of different methods of solution of the partial differential equations: for instance, for finite elements or spectral methods, the array $\vect{s}$ could include the coefficients of the basis functions. $\matr{A}$ is usually a sparse matrix and for approaches founded on the discretization of integral balance equations, it is also symmetric and definite positive.

Finally, \eref{eq:prototype} should be modified to provide a correct formal description of non-linear processes, when $\matr{A}$ and $\vect{b}$ depend on $\vect{s}$.

The data array includes both the positions of the points where potential is measured and the measured values. It could include also other measurements necessary to estimate the values of the source terms.

This example, which refers to a non-linear model, leaves more choices open for the definition of the calibration target and of the model predictions, as will be discussed in the successive sections.

\subsection{Generalities on IP}\label{sec:GeneralitiesOnIp}

With the notation and the definitions given in Sect.~\ref{sec:BasicDefinitions}, the IP should aim to solve the system
\begin{equation}\label{eq:IdealIP}
	\vect{y}\left(\vect{s},\vect{p},\vect{d}\right) = \vect{t}\left(\vect{d},\vect{p}^\mathrm{(fix)}\right),
\end{equation}
with respect to $\vect{p}^\mathrm{(cal)}$. However, an exact solution to \eref{eq:IdealIP} can be rarely found.

A case where an exact solution to \eref{eq:IdealIP} can be found, known in the literature on groundwater hydrology as ``direct approach'' \cite{Neuman1973}, happens if the IP can be cast in the direct formulation, which can be discussed by making reference to the paradigmatic example 2. The model forecast is the equation (or balance) error, namely $\matr{A}\left( \vect{p} \right) \vect{s} - \vect{b}\left( \vect{p} \right)$, i.e. the left hand side of \eref{eq:prototype_f}, where $\vect{s}$ is substituted by values interpolated from the data. In this case, the whole process from data collection to parameter identification can be taken into account with the framework proposed in this paper, if  $\vect{p}^\mathrm{(fix)}$ includes the parameters used for the interpolation of field data. Then, the interpolated state of the system can be expressed as $\vect{s}^\mathrm{(int)}\left(\vect{d},\vect{p}^\mathrm{(fix)}\right)$. On the other hand, the calibration target is $\vect{t}=0$. Then, the direct formulation of the IP reduces to finding the parameters $\vect{p}^\mathrm{(cal)}$ that satisfy
\begin{equation}\label{eq:DirectIP}
	\vect{f}\left(\vect{p},\vect{s}^\mathrm{(int)}\left(\vect{d},\vect{p}^\mathrm{(fix)}\right)\right) = 0.
\end{equation}

In general, it is impossible to guarantee the existence of a solution to \eref{eq:DirectIP}, unless very restrictive conditions are given. However, one can rely on the hypothesis that all the approximations introduced in the model and the quality of the data and of the processing tools are correct enough, so that there exists a set of parameters for which \eref{eq:DirectIP} is satisfied.

In order to overcome this difficulty, the most common approach is to cast the IP in the framework of optimal control and to look for the set of model parameters that minimizes the discrepancy between the two sides of \eref{eq:IdealIP}. This is done by introducing an objective function $\mathsf{O}$, given by:
\begin{equation}\label{eq:ObjectiveFunction}
\mathsf{O} \left( \vect{p}^\mathrm{(cal)} \right) = \mathsf{d} \left( \vect{y}\left(\vect{s},\vect{p},\vect{d}\right), \vect{t}\left(\vect{d},\vect{p}^\mathrm{(fix)}\right) \right),
\end{equation}
where
\begin{itemize}
	\item $\mathsf{d} : \mathbb{R}^{N_\mathrm{(t)}}\times\mathbb{R}^{N_\mathrm{(t)}} \to [0,+\infty)$,
	\item $\vect{y} = \vect{t}\quad \Longleftrightarrow\quad \mathsf{d} \left( \vect{y}, \vect{t} \right) = 0$.
\end{itemize}

The classical choice is the least-squares approach, for which 
\begin{equation}\label{eq:dforLS}
\mathsf{d} \left( \vect{y}\left(\vect{s},\vect{p},\vect{d}\right), \vect{t}\left(\vect{d},\vect{p}^\mathrm{(fix)}\right) \right) = \sum_{i=1}^{N_{\mathrm{(t)}}} \left| \vect{y}_i\left(\vect{s},\vect{p},\vect{d}\right) - \vect{t}_i\left(\vect{d},\vect{p}^\mathrm{(fix)}\right) \right|^2.
\end{equation}

Of course many other choices are possible for the $\mathsf{d}$ function, among which the sum of absolute differences ($\ell^1$ norm) and the maximum absolute difference ($\ell^\infty$ norm), or any $\ell^p$ norm of the difference $\vect{y}\left(\vect{s},\vect{p},\vect{d}\right) - \vect{t}\left(\vect{d},\vect{p}^\mathrm{(fix)}\right)$.

Therefore, one can give the following general definition.
\begin{definition}[Inverse problem -- IP] Let $\vect{s}=\vect{g}\left(\vect{p}\right)$ be the solution to the FP from \eref{eq:ExplicitModel} (indirect approach) or let $\vect{s}=\vect{s}^{(\mathrm{int})}\left(\vect{d}, \vect{p}^{(\mathrm{fix})}\right)$ be the interpolated state of the system over the discretization grid (direct approach). Given $\vect{p}^\mathrm{(fix)}$ and $\vect{d}$, given $$\mathcal{P}^{(\mathrm{cal})}=\left\{ \vect{p}^{(\mathrm{cal})}: \left( {\vect{p}^\mathrm{(fix)}}^t, {\vect{p}^\mathrm{(cal)}}^t \right)^t \in \mathcal{P} \right\},$$ given the functions $\vect{y}$ and $\vect{t}$ -- in particular, the values of $\vect{t}\left(\vect{d},\vect{p}^\mathrm{(fix)}\right)$ --, and given the objective function $\mathsf{O}$ from \eref{eq:ObjectiveFunction}, the IP is finding ${\vect{p}^{(\mathrm{cal})}}^\star \in\mathcal{P}^{(\mathrm{cal})}$, such that
\begin{equation}\label{eq:DefIP}
	\begin{array}{l}
	\mathsf{O} \left( {\vect{p}^{(\mathrm{cal})}}^\star \right) \le \mathsf{O} \left( \vect{p}^\mathrm{(cal)} \right),\ \forall \vect{p}^\mathrm{(cal)}\in\mathcal{P}^{(\mathrm{cal})}\\
	\mbox{or } \displaystyle {\vect{p}^{(\mathrm{cal})}}^\star = \arg\min_{\vect{p}^\mathrm{(cal)}\in\mathcal{P}^{(\mathrm{cal})}} \mathsf{O} \left( \vect{p}^\mathrm{(cal)} \right).
	\end{array}
\end{equation}
\end{definition}

Notice that if $\vect{y}=\vect{s}$ and $\vect{t}\subset\vect{d}$ for a linear model, as discussed for example 1, the application of the least-squares approach simply reduces to solving the linear system of equations
\begin{equation}\label{eq:LinearLeastSquares}
	\matr{G}^t\matr{G}\vect{p}^\mathrm{(cal)}=\matr{G}^t\vect{t},
\end{equation}
so that the properties of the solution to the IP depend on the properties of the $\matr{G}^t\matr{G}$ matrix.

Another remark is relevant with respect to the $\mathsf{d}$ function of \eref{eq:ObjectiveFunction}. The conditions which have been introduced for this function are very similar to those which define a distance or metric among the elements of a given set. However, this function does not necessarily correspond to a distance, as two properties which define a distance are not required here, namely symmetry and triangle inequality. As an example, it is important to recall that even some commercial codes use the root mean square relative error, given by
\begin{equation}\label{eq:RMSRelErr}
\mathsf{d} \left( \vect{y}, \vect{t} \right) = \left[ \frac{1}{N_{\mathrm{(t)}}} \sum_{i=1}^{N_{\mathrm{(t)}}} \left| \frac{\vect{y}_i - \vect{t}_i}{\vect{t}_i} \right|^2 \right]^{1/2}.
\end{equation}
The idea behind this choice is to weight the discrepancy between model forecast and calibration target by considering the magnitude of the target. In other words, a small discrepancy could be negligible if the target value should be great, whereas it is of great relevance if the target value is small, of the same order of magnitude. Under this premise, the symmetry condition is not physically significant, i.e., it is physically very different to normalize with respect to $\vect{t}$ or to $\vect{y}$. And this impacts also on the asymmetry of the function.

Nevertheless, even if one disregards this fact, it is easy to check that the function $\mathsf{d}$ defined by \eref{eq:RMSRelErr} does not satisfy the triangle inequality. For this, it suffices to consider $N_\mathrm{(t)}=1$ and the three numbers $a=1$, $b=4$ and $c=2$ to see that $3 = \mathsf{d}(a,b) > \mathsf{d}(a,c) + \mathsf{d}(c,d) = 2$, which violates the triangle inequality.

\section{Properties related to the IP}

The definitions given in the previous section are now used to introduce several properties of the IP. The first key question is whether the model parameters are identifiable, or, in other words, if different values of the parameters always yield different predictions of the state of the system with the FP. Then the well- or ill-posedness of the IP will be discussed.

\subsection{Identifiability}\label{sec:Identifiability}
Identifiability is a property of the FP, but is strictly related to the IP, as it will be shown in Sec.~\ref{sec:uniqueness}. It is defined as follows \cite{Giudici1989,Giudici1991,KitamuraNakagiri,SunYeh1990}.
\begin{definition}[Identifiability] The model parameters are said to be identifiable, if different sets of model parameters yield different solutions to \eref{eq:DiscreteModel}, i.e., if for every couple of arrays $\vect{p}$ and $\vect{p}'$, $\vect{p}\ne\vect{p}'$, the corresponding solutions to \eref{eq:DiscreteModel}, $\vect{s}=\vect{g}\left( \vect{p} \right)$ and $\vect{s}'=\vect{g}\left( \vect{p}' \right)$ are such that $\vect{s}\ne\vect{s}'$.
\end{definition}

Since identifiability is a very general and strong condition, a weaker formulation can be given as follows.
\begin{definition}[Conditional identifiability] The model parameters are said to be conditionally identifiable if the condition of identifiability holds for a given subspace of state arrays, i.e., if it holds only when $\vect{s}$ and $\vect{s}'$ belong to a given subspace $\mathcal{S}^\mathrm{(c)}\subseteq\mathcal{S}$.
\end{definition}

This permits to define the identifiability of a single parameter as follows.
\begin{definition}[Identifiability of a single parameter] A model parameter $\vect{p}_i$ is said to be identifiable if the condition of identifiability holds for all the couples of arrays $\vect{p}$ and $\vect{p}'$, which differ from each other only for the $i$-th parameter, and for a subspace of state arrays, in other words when $\vect{s}$ and $\vect{s}'$ belong to a subspace $\mathcal{S}^\mathrm{(c)}_i\subseteq\mathcal{S}$.
\end{definition}
For instance, for the paradigmatic example 2, $a_{i+1/2}$ is identifiable if $\vect{s}_i\neq \vect{s}_{i+1}$, i.e., if the discrete gradient of the state of the system between the nodes $i$ and $i+1$ is not null: such a condition can be stated as $\vect{s}\in\mathcal{S}^\mathrm{(c)}_i=\left\{ \vect{s} : \vect{s}_i\neq \vect{s}_{i+1} \right\}$.

The notion of identifiability is of great importance for the direct approach to the IP, as it is equivalent to the uniqueness of the inverse mapping. However, it is related also to the uniqueness of the indirect approach to the IP, as shown in the next subsection.

\subsection{Well-posedness}\label{sec:well-posedness}
Any mathematical problem that is relevant to simulate physical processes is expected to be well-posed. However, it is well known that IP theory is the ``natural habitat'' of ill-posed problems. Comments on the existence of a solution have already been given in section \ref{sec:GeneralitiesOnIp}. It is now time to discuss uniqueness and stability.

\subsubsection{Uniqueness}\label{sec:uniqueness}
In principle, it is very easy to define uniqueness of the IP as the property that a unique array ${\vect{p}^{(\mathrm{cal})}}^\star$ satisfies \eref{eq:DefIP}. In principle, it is also very easy to state that a sufficient condition for IP to admit a unique solution is that $\mathsf{O}$ be a convex strictly function, i.e.
\begin{equation}\label{eq:Convexity}
\begin{array}{l}\mathsf{O} \left( \lambda \vect{p} + (1-\lambda) \vect{p}' \right) < 
\lambda \mathsf{O} \left( \vect{p} \right) + (1-\lambda) \mathsf{O} \left( \vect{p}' \right),\\ \forall \lambda\in [0,1],\ \forall \vect{p},\vect{p}'\in\mathcal{P}^{(\mathrm{cal})},\ \vect{p}\neq\vect{p}'.
\end{array}
\end{equation}
Unfortunately, it is not easy to check if \eref{eq:Convexity} holds or to prove theorems that give necessary and sufficient conditions for its validity that are practically useful.

Obviously, if one of the model parameters that belong to $\vect{p}^{(\mathrm{cal})}$, say $\vect{p}_i$, is not identifiable, neither conditionally identifiable, if a solution $\vect{p}^\dagger$ to \eref{eq:DefIP} is such that $\vect{g}\left( \vect{p}^\dagger \right)\notin \mathcal{S}^\mathrm{(c)}_i$, 
then there exists $\vect{p}^\ddagger$ such that $\vect{g}\left( \vect{p}^\ddagger \right) = \vect{g}\left( \vect{p}^\dagger \right)$. If $\vect{y}$ does not explicitly depend on $\vect{p}_i$, as it is often the case, then $\vect{y}\left( \vect{d},\vect{g}\left( \vect{p}^\dagger \right),\vect{p}^\dagger \right) = \vect{y}\left( \vect{d},\vect{g}\left( \vect{p}^\ddagger \right),\vect{p}^\ddagger \right)$. Therefore, if the previous conditions are met, $\mathsf{O} \left( \vect{p}^\dagger \right) = \mathsf{O} \left( \vect{p}^\ddagger \right) $, which proves that the solution to the IP is not unique.

This remark shows the strict link between identifiability, which is a property of the FP, and uniqueness of the IP, even in the indirect formulation.

Notice that for the simple case of a linear model, \eref{eq:LinearLeastSquares} can be solved if 
\begin{equation}\label{eq:detGtG}
\det\left( \matr{G}^t\matr{G} \right)\neq 0.
\end{equation}
In that case, the solution is unique and depends on the input data in a stable way.

For the paradigmatic example 1, the computation of $\matr{G}^t\matr{G}$ from \eref{eq:GEx1} yields:
\begin{equation}\label{eq:GtGEx1}
\matr{G}^t\matr{G} = \frac{L^2}{4} \cdot \left( 
\begin{array}{cccc}
	3 & 1 & 0 & 2\\
	1 & 3 & 2 & 0\\
	0 & 2 & 3 & 1\\
	2 & 0 & 1 & 3
\end{array}
\right),
\end{equation}
which is apparently not invertible, as the first raw is a linear combination of the remaining three rows. When this happens, the IP is said to be underdetermined, as the data are insufficient to determine the values of the model parameters. If a solution $\vect{p}^\dagger$ of an underdetermined IP can be found for a linear model, then infinite solutions can be found as $\vect{p}^\dagger + \vect{p}^{(0)}$, where $\vect{p}^{(0)}$ is an arbitrary element of the null space $\mathcal{N}=\left\{ \vect{p}: \matr{G}^t\matr{G}\vect{p}=0 \right\}$. For example 2, the null space is given by elements $\vect{p}^{(0)}=(c,-c,-c,c)^t$, where $c\in\mathbb{R}$ is arbitrary.

The issue of uniqueness depends clearly on the data. For example, it has been shown \cite{GIN09} that a high sampling density of piezometric data does not prevent non-uniqueness of the hydraulic conductivity; on the other hand, non-uniqueness is reduced when groundwater age data are considered together with piezometric data in the IP. Also joint inversion of hydraulic head and solute concentration is helpful to establish uniqueness \cite{KNOWLES2004277}.

\subsubsection{Stability and conditioning}\label{sec:StabilityAndConditioning}
The inverse problem is usually claimed to be unstable. In fact, very simple examples show that this is true for the IP in the continuous case, for instance when dealing with example 2 in a continuous domain so that the physics is expressed through partial differential equations \cite{GiudiciVassena2008}.

In order to cast the problems in a precise way, it is necessary to give the following definition.
\begin{definition}[Stability] If $\vect{d}^\dagger$ and $\vect{d}^\ddagger$ are two sets of data and $\vect{p}^\dagger$ and $\vect{p}^\ddagger$ are the corresponding solutions to \eref{eq:DefIP}, the IP is stable if
\begin{equation}\label{eq:stability}
\left\|\vect{p}^\dagger - \vect{p}^\ddagger \right\| \to 0
\mbox{ as }
\left\| \vect{d}^\dagger - \vect{d}^\ddagger \right\| \to 0.
\end{equation}
\end{definition}

The big difference with respect to the case of the IP for a continuous medium is that the solution of the discrete IP is essentially based on a sequence of algebraic operations, which could be stable. However, stability is a mathematical property, which assumes that the error on the data can be reduced at will, so that also the calibrated parameters converge to the ``correct'' values. Unfortunately, the difference between model predictions and calibration targets depends on several sources: the accuracy of the measuring instruments; the correctness of the acquisition procedures; the relevance of the measurement support volumes with respect to the spatial and temporal scales of the model; the model approximations; the spatial and temporal discretization grid; etc. Some of these factors cannot be reduced in practice, because they are fixed when a model is applied. Therefore, even if stability has a fundamental importance from the mathematical point of view, well-conditioning of the IP is even more important from the physical point of view \cite{Giudici2002,Aster2013}. Conditioning is defined as follows.
\begin{definition}[Conditioning] Under the same hypotheses given for the definition of stability and if the following Lipschitz's condition is satisfied
\begin{equation}\label{eq:condition}
\left\|\vect{p}^\dagger - \vect{p}^\ddagger \right\| 
\le C \left\| \vect{d}^\dagger - \vect{d}^\ddagger \right\|,
\end{equation}
where $C$ is a constant value, the IP is said to be well-conditioned if $C$ is small and ill-conditioned if $C$ is big.
\end{definition}
This definition is rather qualitative, but it is of great value in order to properly assess the physical relevance of the calibrated model parameters. In fact, for an ill-conditioned problem, the error on the calibrated parameters could be very high because of the enhancing Lipschitz factor $C$, even if the error on the data is small; on the other hand, for a well-conditioned IP, larger errors on the input data could nevertheless yield acceptable values of the calibrated model parameters because of the low value of $C$.

Notice that even for the simplest case of an IP for a linear model, the condition \eref{eq:detGtG} guarantees that the IP is stable, but the IP could be ill-conditioned. A classical example from linear algebra is the matrix
\begin{equation}
\matr{A} = \left( \begin{array}{cc}
\epsilon & 0 \\
0 & 1/\epsilon
\end{array} \right),
\end{equation}
where $\epsilon$ is a small quantity, for which $\det\matr{A}=1$, so that $\matr{A}^{-1}$ can be computed in a unique and stable way, but the condition number, computed as the ratio between the maximum and minimum eigenvalues of the matrix, is given by $\epsilon^{-2}$, which is a great number. The practical consequence is that if the elements of the matrix are estimated with a small error, of the same order of magnitude as $\epsilon$, then the computation of the inverse matrix might be affected by unacceptably large errors.

\begin{figure}[htbp]
	\centering
		\includegraphics[width=10cm]{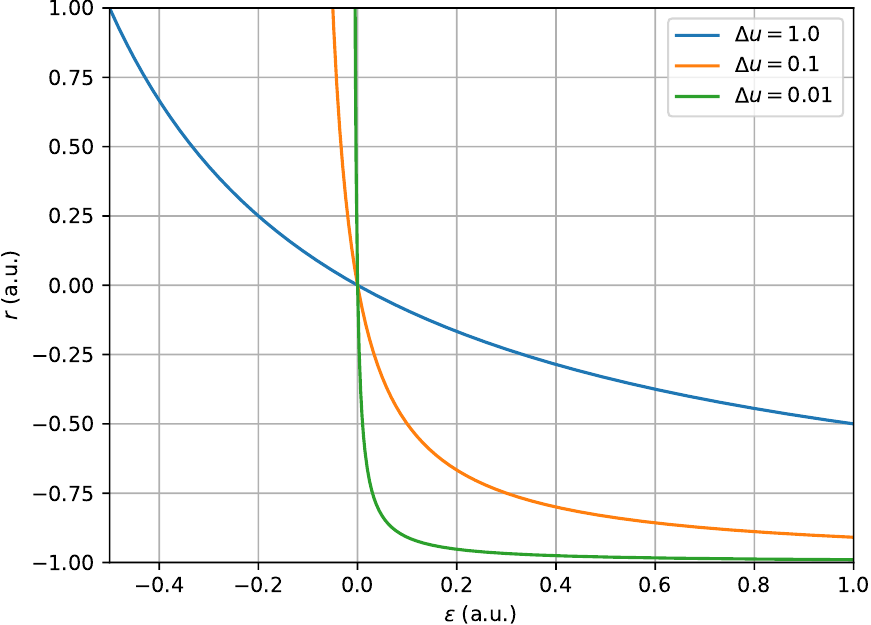}
	\caption{Relative error on the model parameters for a simple case of the paradigmatic example 2, as a function of the error (in arbitrary units) on the estimate of a state variable for three values of $\Delta u$ (green line for 0.01 a.u.; orange line for 0.1 a.u.; cyan line for 1 a.u.).}
	\label{fig:Figure03}
\end{figure}

The concept of conditioning for IP can be conveniently discussed with reference to the paradigmatic example 2. Let us consider a very simple case, when $N=3$ and the grid spacing is uniform and equal to $\Delta x$, so that \eref{eq:FiniteDifference} reduces to
\begin{equation}\label{eq:FiniteDifferenceForN=3}
\begin{array}{lcl}
\displaystyle a_{1/2}\frac{u_{0}-u_{1}}{\Delta x} + a_{3/2}\frac{u_{2}-u_{1}}{\Delta x} &=& \varphi_1,\\
\displaystyle a_{3/2}\frac{u_{1}-u_{2}}{\Delta x} + a_{5/2}\frac{u_{3}-u_{2}}{\Delta x} &=& \varphi_2.
\end{array}
\end{equation}
Let us assume that $\varphi_1$, $\varphi_2$ and $q_{1/2}=a_{1/2}\frac{u_{0}-u_{1}}{\Delta x}$ can be estimated from the data and the array $\vect{p}^\mathrm{(cal)}$ reduces only to two elements: $\vect{p}^\mathrm{(cal)} = \left(a_{3/2}, a_{5/2}\right)^t$. In this case the solution of the IP is unique and given by:
\begin{equation}\label{eq:SolutionIPSimple}
\begin{array}{lcl}
a_{3/2} &=& \left( \varphi_1 - q_{1/2} \right)\cdot \displaystyle \frac{\Delta x}{u_{2}-u_{1}},\\
a_{5/2} &=& \left( \varphi_2 + \varphi_1 - q_{1/2} \right)\cdot \displaystyle \frac{\Delta x}{u_{3}-u_{2}}.
\end{array}
\end{equation}
Let us denote with a hat the quantities estimated or obtained from noise-free data and let us explicitly consider an error on $u_2$, so that the estimated value is given by $\hat{u}_2+\varepsilon$. Then from \eref{eq:SolutionIPSimple}, after simple manipulations, it follows
\begin{equation}\label{eq:SolutionIPSimpleNoisy}
\begin{array}{lcl}
a_{3/2}(\varepsilon) &=& \hat a_{3/2} \cdot \left( 1 + \displaystyle \frac{\varepsilon}{\hat u_2 - \hat u_1} \right)^{-1},\\
\\
a_{5/2}(\varepsilon) &=& \hat a_{5/2} \cdot \left( 1 - \displaystyle \frac{\varepsilon}{\hat u_3 - \hat u_2} \right)^{-1}.
\end{array}
\end{equation}
From \eref{eq:SolutionIPSimpleNoisy} the error on the estimated values of the model parameter is easily expressed as
\begin{equation}\label{eq:SolutionIPSimpleError}
\begin{array}{lcl}
a_{3/2}(\varepsilon) - \hat a_{3/2} &=& \hat a_{3/2} \cdot \left[ \left( 1 + \displaystyle \frac{\varepsilon}{\hat u_2 - \hat u_1} \right)^{-1} - 1 \right],\\
\\
a_{5/2}(\varepsilon) - \hat a_{5/2} &=& \hat a_{5/2} \cdot \left[ \left( 1 - \displaystyle \frac{\varepsilon}{\hat u_3 - \hat u_2} \right)^{-1} - 1 \right].
\end{array}
\end{equation}
Now, a few remarks about \eref{eq:SolutionIPSimpleError}. The errors on $a_{3/2}$ and $a_{5/2}$ depend on the function $r(\varepsilon;\Delta u) = \left( 1 + \varepsilon/\Delta u \right)^{-1} - 1$, where $\Delta u = \hat u_2 - \hat u_1$ and $\Delta u = \hat u_2 - \hat u_3$, respectively for $a_{3/2}$ and $a_{5/2}$. The function $r(\varepsilon)$ represents the relative error on the two calibrated parameters as a function of the error on $u_2$ and its graph is shown in Fig.~\ref{fig:Figure03} for three different values of $\Delta u$ (0.01, 0.1 and 1, expressed with the same arbitrary units as $\varepsilon$). Recall that if $\varepsilon=-\Delta u/2$, $r(\varepsilon)=+1$ and the calibrated parameter attains a value twice the ``true'' one; moreover, if $\Delta u>0$ and $\varepsilon<-\Delta u$, then $r(\varepsilon)<-1$ and the calibrated parameter is negative and therefore not physically acceptable, if interpreted as a ``conductivity'' or ``conductance''. The latter comment explains why the curves in Fig.~\ref{fig:Figure03} are drawn only for $\varepsilon>-\Delta u$. If $\hat u_2 - \hat u_1 = \hat u_3 - \hat u_2$, then the relative errors for the two elements of $\vect{p}^\mathrm{(cal)}$ have opposite sign, so that if one is underestimated, the other is overestimated. The stability and well-conditioning of the IP is controlled by these functions. In particular, it is clear that the IP is stable unless $\Delta u=0$, but in that case one of the parameters is not identifiable, as discussed in section \ref{sec:Identifiability}. However, the IP is well-conditioned if $\Delta u$ is big (see, e.g., the cyan line of Fig.~\ref{fig:Figure03}), since in that case even relatively high values of $\varepsilon$ do not yield great values of $r(\varepsilon)$. On the other hand, when $\Delta u$ is small (see e.g., the green line of Fig.~\ref{fig:Figure03}), even a small value of $\varepsilon$ is sufficient to produce high values of $r(\varepsilon)$ or physically inconsistent values of the calibrated parameters, for instance negative conductances. See also the discussion of this topic in \cite{Giudici2002}.

Note, in particular, that ill-conditioning or discrete instability often appear as an oscillating behaviour of the outcome values of algebraic computations. This is somehow similar to the effect that might arise from non-uniqueness, as shown in section \ref{sec:uniqueness}, where the null space for example 1 has been proved to be composed of arrays whose elements are alternately high and low. These comments motivated the use of regularization for model calibration. Section \ref{sec:regularization} shows how regularization is embedded in the proposed conceptual framework.

\subsection{Properties depending on solution methods}
Section \ref{sec:well-posedness} was devoted to some basic definitions and concepts about the IP. However, it is important to stress that a key factor for a reliable inversion is also the method of solution. In fact, several different algorithms could be applied to find the minimum of $\mathsf{O}$. Many of them are iterative procedures that start from a tentative guess which is progressively improved, often by moving along the steepest-descent direction, i.e. along the gradient of $\mathsf{O}$, or conjugate directions \cite{NumericalRecipes}. These methods are usually very much influenced by the initial guess, which might cause the algorithm to fall into a local minimum, without reaching the global minimum of the function. In order to escape from local minima, it is necessary to apply other algorithms that can span the parameter space $\mathcal{P}$ in a more appropriate way, often with a stochastic approach; among the others, simulated annealing \cite{Kirkpatrick671} and genetic algorithms \cite{mitchell1998introduction} can be recalled.

A slightly different approach is applied by some methods of IP solution in the direct formulation: the Comparison Model Method \cite[CMM]{computation4010002,DeFilippis2016,Giudici2012,Ponzini1988,WRCR:WRCR3134,VASSENA20081105,Vassena2012} and its variation named successive flux estimation \cite{PASQUIER20061934} and the Dual Constraint Method \cite[DCM]{Brouwer2008,Zijl2004,DCM-book}. These methods are based on the use of the solution to the FP for a tentative set of parameters, namely for a tentative transmissivity field for the applications in groundwater hydrology performed so far. The iterative updating of parameters is based on the use of phenomenological laws (Darcy's law in the case of groundwater hydrology). Therefore, these methods have a quite strong physical foundation, but are not based on the use of the gradient of $\mathsf{O}$ to search the minimum of the objective function. They are computationally very fast, but they might suffer from the typical drawbacks related to the use of a direct inversion, mentioned in Sec.~\ref{sec:GeneralitiesOnIp}.

These short remarks suggest that the chosen method of solution could dramatically modify the posedness and the conditioning of the IP. Therefore, as a general comment, it is necessary to clearly distinguish in practical applications the effects of the solution method and the intrinsic properties of the IP.

\section{Some remarks on IP within the proposed conceptual framework}

The proposed conceptual framework provides a new perspective to define and analyse some approaches commonly used in IP theory and applications. Therefore, in the following sections, it is shown how regularization, Bayesian approaches, Kalman filters, multi-objective inversion, sensitivity analysis and the adjoint method can be tackled in more natural and straightforward terms when cast within the proposed conceptual framework.

\subsection{Regularization}\label{sec:regularization}
The remarks of section \ref{sec:StabilityAndConditioning} show that the effect of instability and of non-uniqueness could be the alternation between contrasting (high and low) values for the model parameters. In order to reduce this effect, regularization approaches ``filter'' such an oscillating behaviour by including a term proportional to the norm of the model parameters in the objective function. This approach is easily embedded in the developed conceptual framework, by including sub-arrays $\vect{y}^{(\mathrm{reg})}$ and $\vect{t}^{(\mathrm{reg})}$ in the arrays $\vect{y}$ and $\vect{t}$. The simplest strategy for this goal is setting $\vect{y}^{(\mathrm{reg})}=\vect{p}^{(\mathrm{cal})}$ and $\vect{t}^{(\mathrm{reg})}=0$. If the $\ell^2$ norm is used to define $\mathsf{O}$, then the above definitions yield a contribution $\sum_{i=1}^{N_\mathrm{(p)}} \left| \vect{p}^{(\mathrm{cal})}_i \right|^2$, as required.

\subsection{Bayesian approach}
The discussion of section \ref{sec:StabilityAndConditioning} emphasizes the role of the measurement, approximation and modelling errors on the solution of IP. The number and complexity of factors that affect those errors prompted researchers to use statistics and the theory of stochastic processes in order to properly account for errors in the IP solution. The common strategy invokes the Bayes' theorem, which can be cast, with the formalism of the proposed framework, as:
\begin{equation}\label{eq:BayesTheorem}
	f \left( \vect{p}^\mathrm{(cal)} | \vect{y} - \vect{t} \right) = \frac{ f \left( \vect{y} - \vect{t} | \vect{p}^\mathrm{(cal)} \right) \cdot f \left( \vect{p}^\mathrm{(cal)} \right) }{ f \left( \vect{y} - \vect{t} \right) },
\end{equation}
where $f$ functions are (possibly conditioned) probability density functions (pdfs) of the respective arguments. In particular, $f \left( \vect{p}^\mathrm{(cal)} \right)$ is the prior pdf of the model parameters to be calibrated and is independent from the measurements of state variables or other independent quantities; instead, $f \left( \vect{p}^\mathrm{(cal)} | \vect{y} - \vect{t} \right)$ represents the posterior pdf, which is conditioned on the measured data. Notice that \eref{eq:BayesTheorem} is slightly different from the standard formulation proposed by other Authors, who consider $\vect{y} - \vect{t} = \vect{s} - \vect{d}$. In fact, the framework introduced in this paper is more general, because it allows to include different types of model outputs and calibration targets.

The Bayesian approach is usually coupled with the maximum likelihood method \cite{Edwards1972}, which states the IP as follows.
\begin{definition}[Maximum likelihood] Under the same hypotheses introduced to define the IP solution and with the same notation used for Bayes' theorem, the maximum likelihood solution to the IP is given by the array ${\vect{p}^{(\mathrm{cal})}}^\ast$ for which the likelihood function 
\begin{equation}\label{eq:Likelihood}
\mathcal{L} \left( \vect{p}^\mathrm{(cal)} \right) = f \left( \vect{p}^\mathrm{(cal)} | \vect{y} - \vect{t} \right),
\end{equation}
or a monotonically increasing transform of $\mathcal{L} \left( \vect{p}^\mathrm{(cal)} \right)$, attains its maximum value.
\end{definition}

The most useful transform of $\mathcal{L}$ is a logarithmic function, which, together with \eref{eq:BayesTheorem}, makes it possible to rewrite the function to be maximized as
\begin{equation}\label{eq:LogL}
	\ln\mathcal{L} \left( \vect{p}^\mathrm{(cal)} \right) = \ln \left[ f \left( \vect{y} - \vect{t} | \vect{p}^\mathrm{(cal)} \right) \right] + \ln \left[ f \left( \vect{p}^\mathrm{(cal)} \right) \right] - \ln \left[ f \left( \vect{y} - \vect{t} \right) \right].
\end{equation}
Notice that in most cases the last term of the right hand side of \eref{eq:LogL} is not considered, because it is often defined in such a way as to be independent of $\vect{p}^\mathrm{(cal)}$. Also, the Bayesian approach implicitly introduces a regularization through the prior distribution $f \left( \vect{p}^\mathrm{(cal)} \right)$.

The application of this approach obviously requires some guesses about the pdfs appearing in \eref{eq:LogL}. The most common guess consists in assuming that both $f \left( \vect{y} - \vect{t} | \vect{p}^\mathrm{(cal)} \right)$ and $f \left( \vect{p}^\mathrm{(cal)} \right)$ can be expressed as multi-Gaussian distributions. Under these hypotheses, if the last term of the right hand side of \eref{eq:LogL} is neglected, the maximum-likelihood method reduces to the least-squares approach. More precisely, it reduces to the weighted least-squares, where the differences $\vect{y} - \vect{t}$ and the discrepancy of $\vect{p}^\mathrm{(cal)}$ from its prior expected-value are weighted by the inverse of the corresponding covariance matrices. Such matrices take a diagonal form if the differences $\vect{y}_i - \vect{t}_i$ are independent of each other; moreover, if the variances of all the differences are independent from $i$, then the maximum-likelihood method reduces to the standard least-squares method.

However, this is not the only possible choice. In fact, a multi-Gaussian distribution is not the best option if outliers are present, i.e., if some values of $\vect{y}$ are very far from the expected target value $\vect{t}$. In those cases, an exponential distribution might provide a better approximation of the pdfs, in particular of $f \left( \vect{y} - \vect{t} | \vect{p}^\mathrm{(cal)} \right)$, and it practically leads to the minimization of $\ell^1$-like norms.

The above remarks are fundamental, because they implicitly show which are the conditions and the physical requirements at the basis of the classical least-squares approach.

\subsection{Kalman filter approach}

Non-linear estimation can be obtained also with the application of the Ensemble Kalman filter (EnKF). The standard Kalman filter was developed for estimating the system state predicted by a linear model, by profiting from direct measurements. The EnKF is an extension and modification of the Kalman filter, in order to work with non-linear models and when the initial states are uncertain \cite{JGRC:JGRC5955,Evensen2003}. It can also be applied to perform the identification of model parameters with the procedure that is briefly recalled in this section.

Let data corresponding to different time steps, denoted with the index
$k$, be available, together with an evolutionary model. Then
calibration targets at different times can be collected in arrays
$\vect{t}^{(k)}$ and a sequence of steps is performed iteratively:
starting from an initial guess which follows a given pdf, an ensemble
of $N_\mathrm{(r)}$ estimates (prior estimate) of
$\vect{p}^{(\mathrm{cal})}$ is assumed to be available at a given step
$k$; each element of this ensemble is denoted as
${\vect{p}^{(\mathrm{cal})}}^{(k,q)}$, where the index
$q=1,\ldots,N_\mathrm{(r)}$ is used to identify the realizations of
model parameters. The model predictions can be computed for each
realization and are collected in arrays $\vect{y}^{(k,q)}$, whereas
measurements are used to build the arrays of time-dependent
calibration targets $\vect{t}^{(k)}$.

Posterior estimates ${\vect{p}^{(\mathrm{cal})}}^{(k+1,q)}$ of the model parameters are obtained by means of a corrective term, as follows
\begin{equation}\label{eq:KalmanFilter}
{\vect{p}^{(\mathrm{cal})}}^{(k+1,q)} = {\vect{p}^{(\mathrm{cal})}}^{(k,q)} + \matr{K}^{(k)} \left( \vect{y}^{(k,q)} - \vect{t}^{(k)} \right),
\end{equation}
where $\matr{K}^{(k)}$ is called the gain matrix in the standard Kalman filter. Its expression for the EnKF is 
\begin{equation}\label{eq:matK}
\matr{K}^{(k)} = \mathrm{Cov}^{(k)}[\vect{p}\vect{y}] \cdot \mathrm{Cov}^{(k)}[\vect{y}\vect{y}]^{-1},
\end{equation}
where $\mathrm{Cov}^{(k)}[\vect{p}\vect{y}]$ and $\mathrm{Cov}^{(k)}[\vect{y}\vect{y}]$ are covariance matrices, whose elements are computed as
\begin{equation}\label{eq:cov-py}
\begin{array}{l}
\mathrm{Cov}^{(k)}[\vect{p}\vect{y}]_{lj} =\\
= \displaystyle \frac{1}{N_\mathrm{(r)}-1}\sum_{q=1}^{N_\mathrm{(r)}} \left({\vect{p}^{(\mathrm{cal})}}^{(k,q)}_{l} - <{\vect{p}^{(\mathrm{cal})}}^{(k)}_{l}>) \right) \left( {\vect{y}^{(k,q)}}_{j} - <{\vect{y}^{(k)}}_{j}>) \right),\\
l=1,\ldots,N_\mathrm{(c)},\ j=1,\ldots,N_\mathrm{(t)},
\end{array}
\end{equation}
and
\begin{equation}\label{eq:cov-yy}
\begin{array}{l}
\mathrm{Cov}^{(k)}[\vect{y}\vect{y}]_{ij} =\\
= \displaystyle  \frac{1}{N_\mathrm{(r)}-1}\sum_{q=1}^{N_\mathrm{(r)}} \left( {\vect{y}^{(k,q)}}_{i} - <{\vect{y}^{(k)}}_{i}>) \right) \left( {\vect{y}^{(k,q)}}_{j} - <{\vect{y}^{(k)}}_{j}>) \right),\\
i,j=1,\ldots,N_\mathrm{(t)},
\end{array}
\end{equation}
where $<\cdot>$ denotes the ensemble average of its argument.

This procedure is repeated iteratively, so that a succession of ensembles of parameter distributions ${\vect{p}^{(\mathrm{cal})}}^{(k)}$ is obtained, which eventually provides a statistical distribution of calibrated parameters.

\subsection{Multi-objective inversion}
The definition of IP given in section \ref{sec:GeneralitiesOnIp} includes the possibility of using different kinds of physical quantities to be used as calibration targets and, therefore, it implicitly accounts for multi-objective inversion \cite{GJI:GJI977}. The goal of this section is not to discuss and propose new results in multi-objective inversion, but to show how multi-objective inversion concepts can be easily formulated in terms of the proposed conceptual framework. In fact the objective function can be esplicitly defined as the sum of several positive objective functions, $\mathsf{O}^{(k)}\left( \vect{p}^\mathrm{(cal)} \right)$, $k=1,\ldots,N_\mathrm{(O)}$:
\begin{equation}\label{eq:MultiObjectiveInversion}
	\mathsf{O}\left( \vect{p}^\mathrm{(cal)} \right) = \sum_{k=1}^{N_\mathrm{(O)}} \mathsf{O}^{(k)}\left( \vect{p}^\mathrm{(cal)} \right).
\end{equation}

Multi-objective inversion is often applied by considering individual objective functions which yield complementary information and which depend in very different ways on $\vect{p}^\mathrm{(cal)}$. Then one is obliged to admit that there is no optimal solution; in fact, a global minimum could hardly be found and several sets of model parameters could permit to fit the calibration target in a reasonable way, despite not being ``optimal''. This prompted some researchers to import the concept of Pareto optimality or efficiency from quantitative economics to geophysics, in particular to hydrology \cite{baratelli2014,baratelli2011,GUP98,MAD03,VRU03,YAP98}: a set of parameters is said to be Pareto optimal (or Pareto efficient, or non-dominated), if none of the objective functions can be improved in value without degrading some of the other objective values. This can be stated formally as follows.
\begin{definition}[Pareto optimal solution] An array ${\vect{p}^\mathrm{(cal)}}^\dagger$ is said to be a Pareto optimal (or non-dominated or Pareto efficient) solution of the IP for \eref{eq:MultiObjectiveInversion}, if
\begin{equation}\label{eq:ParetoOptimal}
	\nexists\ \mathrm{p}^\mathrm{(cal)}\in\mathcal{P}^{(\mathrm{cal})} \ :\ \mathsf{O}^{(k)}\left( \vect{p}^\mathrm{(cal)} \right) \le \mathsf{O}^{(k)}\left(  {\vect{p}^\mathrm{(cal)}}^\dagger \right),\ \forall k=1,\ldots,N_\mathrm{(O)}.
\end{equation}
The set of Pareto optimal solutions is called the Pareto set or Pareto frontier.
\end{definition}
When the Pareto set reduces to a single array, this corresponds to the minimum of each individual objective function $\mathsf{O}^{(k)}$, $k=1,\ldots,N_\mathrm{(O)}$.

\subsection{Sensitivity analysis}
The proposed conceptual framework is very useful also to discuss sensitivity analysis. Sensitivity analysis permits to quantify the uncertainty on model outputs due to the uncertainty on the input model parameters, but its description is outside the goals of this paper. However, it is possible to show how the developed conceptual model is linked to sensitivity indicators, through several definitions \cite{2434_166184,HIL06}.
\begin{definition}[State sensitivity] State sensitivity, $\matr{S}^\mathrm{(s)}_{mn}$, provides the variability of a state variable $\vect{s}_m$ with respect to a single parameter $\vect{p}_n$, under a linear approximation for small local variations of the parameter:
\begin{equation}\label{eq:ss}
	\matr{S}^\mathrm{(s)}_{mn} = \frac{\partial \vect{s}_m}{\partial \vect{p}_n} = \frac{\partial \vect{g}_m}{\partial \vect{p}_n} \left( \vect{p} \right).
\end{equation}
\end{definition}

For a linear model, $\matr{S}^\mathrm{(s)}$ corresponds to the $\matr{G}$ matrix.

\begin{definition}[Prediction sensitivity] Prediction sensitivity, $\matr{S}^\mathrm{(y)}_{mn}$, provides the variability of a model prediction $\vect{y}_m$ with respect to a single parameter $\vect{p}_n$, under a linear approximation for small local variations of the parameter:
\begin{equation}\label{eq:ps}
	\matr{S}^\mathrm{(y)}_{mn} = \frac{\mathrm{d} \vect{y}_m}{\mathrm{d} \vect{p}_n} = \sum_{k=1}^{N_\mathrm{(s)}} \frac{\partial \vect{y}_m}{\partial \vect{s}_k} \cdot \frac{\partial \vect{s}_k}{\partial \vect{p}_n} + \frac{\partial \vect{y}_m}{\partial \vect{p}_n} = \sum_{k=1}^{N_\mathrm{(s)}} \frac{\partial \vect{y}_m}{\partial \vect{s}_k} \cdot \matr{S}^\mathrm{(s)}_{kn} + \frac{\partial \vect{y}_m}{\partial \vect{p}_n}.
\end{equation}
\end{definition}

Notice that in \eref{eq:ps} the ``total'' dependence of $\vect{y}$ on $\vect{p}$ is considered explicitly, and includes both the direct functional dependence and the indirect dependence through the solution of the FP.

If parameters and state systems are physical quantities, with given measurement units, it is impossible to identify the most sensitive parameters from a straightforward comparison among the elements of $\matr{S}^\mathrm{(s)}$ or $\matr{S}^\mathrm{(y)}$. In fact, it is necessary to scale or normalize these quantities. This can be done by scaling both the independent variables (the parameters) and the dependent variables (system state for $\matr{S}^\mathrm{(s)}$ and model predictions for $\matr{S}^\mathrm{(y)}$) with the reference values around which the sensitivity indices are computed; the scaled sensitivities provide the relative variations of $\vect{s}$ and $\vect{y}$ with respect to a unit relative variation of $\vect{p}$. An alternative is normalizing $\matr{S}^\mathrm{(s)}$ and $\matr{S}^\mathrm{(y)}$ with measures of variability, for instance, the standard deviation of the relevant quantities.

Both $\matr{S}^\mathrm{(s)}$ and $\matr{S}^\mathrm{(y)}$ are locally defined quantities and are based on a one-at-a-time approach, so that they take into account only the linear approximation of the model and neglect both non-linear effects and joint effects of the parameters. This can be overcome by considering the input parameters, and therefore the model predictions, as stochastic quantities and by giving the following definition \cite{Sal08}.

\begin{definition}[First-order sensitivity] If $Y$ represents a state variable $\vect{s}_m$ or a model prediction $\vect{y}_m$, then the first-order sensitivity of $Y$ with respect to $\vect{p}_n$ is given by:
\begin{equation}\label{eq:1st-order-sensitivity}
  S_n = \frac{\mathrm{var}_{\vect{p}_n}[E_{\vect{p}\backslash n}[Y|\vect{p}_n]]}{\sigma_Y^2},
\end{equation}
where $E_{\vect{p}\backslash n}[Y|\vect{p}_n]$ is the expected value of $Y$ conditioned on the parameter $\vect{p}_n$ and $\mathrm{var}_{\vect{p}_n}$ is the variance with respect to $\vect{p}_n$.
\end{definition}

\subsection{Adjoint method for the computation of sensitivity}
The computation of $\matr{S}^\mathrm{(s)}$ is often a crucial aspect for the application of IPs. It is necessary to compute not only $\matr{S}^\mathrm{(y)}$, but also the gradient of $\mathsf{O}$ for methods of solution which are based on steepest-descent or conjugate-gradient approaches.

For IPs related to models based on the numerical solution of partial differential equations, like the paradigmatic example 2, $\matr{S}^\mathrm{(s)}$ is often computed by means of the so-called adjoint method \cite{2006GeoJI.167..495P}. This method is often introduced in the continuous case, by making use of variational calculus and by introducing the Frechet's derivative. In this work, instead, the adjoint method is reviewed for a quite wide class of discrete problems, namely those which can be represented by \eref{eq:prototype}.

If \eref{eq:prototype} is multiplied by an arbitrary array $\vect{v}^{(m)}$, and the derivative of the resulting equation with respect to $\vect{p}_n$ is taken, one obtains
\begin{equation}\label{eq:PrototypeTimesPsi}
	\frac{\partial \matr{A}}{\partial \vect{p}_n} \vect{s} \cdot \vect{v}^{(m)} + \matr{A}\frac{\partial \vect{s}}{\partial \vect{p}_n} \cdot \vect{v}^{(m)} - \frac{\partial \vect{b}}{\partial \vect{p}_n} \cdot \vect{v}^{(m)} = 0.
\end{equation}
Then
\begin{equation}\label{eq:DyDs}
  \begin{array}{lcl}
	\displaystyle \frac{\partial \vect{s}_m}{\partial \vect{p}_n} &=& \displaystyle \frac{\partial \vect{s}_m}{\partial \vect{p}_n} - \frac{\partial \matr{A}}{\partial \vect{p}_n} \vect{s} \cdot \vect{v}^{(m)} - \matr{A}\frac{\partial \vect{s}}{\partial \vect{p}_n} \cdot \vect{v}^{(m)} + \frac{\partial \vect{b}}{\partial \vect{p}_n} \cdot \vect{v}^{(m)}\\
	\\
	&=& \displaystyle - \frac{\partial \matr{A}}{\partial \vect{p}_n} \vect{s} \cdot \vect{v}^{(m)} + \frac{\partial \vect{b}}{\partial \vect{p}_n} \cdot \vect{v}^{(m)},
	\end{array}
\end{equation}
provided $\vect{v}^{(m)}$ is the solution of the so-called ``adjoint-state equation''
\begin{equation}\label{eq:AdjointState}
  \matr{A}^t \vect{v}^{(m)} = \bm{\delta}_m,
\end{equation}
where $\bm{\delta}_m$ is the unit impulse concentrated on the $m$-th element. Recall that in most cases $\matr{A}^t = \matr{A}$.

In other words, computing $\matr{S}^\mathrm{(s)}$ with the adjoint-state approach requires the solution of \eref{eq:AdjointState} for each $m$ and then the application of \eref{eq:DyDs}. This procedure could appear cumbersome, but it must be recalled that for the application of the model, it is necessary to have an efficient code, function or routine for the solution of the FP: for a single value of $m$, only one run of the same tool can be used to compute $\vect{v}^{(m)}$ as the solution to \eref{eq:AdjointState}, and then $\matr{S}^\mathrm{(s)}_{mn}$, for $n=1,\ldots,N_\mathrm{(p)}$ by means of \eref{eq:DyDs}. By comparison, the computation of $\matr{S}^\mathrm{(s)}_{mn}$ with a finite-difference approach would require the solution of the FP for two different arrays $\vect{p}^+$ and $\vect{p}^-$, which differ from each other only for the value of $\vect{p}_n$ by an amount $\Delta p$ and which yield solutions to the FP, respectively, $\vect{s}^+$ and $\vect{s}^-$. Then $\left( \vect{s}^+ - \vect{s}^- \right)/\Delta p$ could be used to approximate the searched derivatives. Notice that this simple approach is nevertheless approximate and requires the solution to FPs. The adjoint-state approach, based on \eref{eq:AdjointState} and \eref{eq:DyDs}, yields a result, which is theoretically perfect and affected only by rounding errors.

Moreover, if the values of the system state are measured, so that $\vect{y}=\vect{s}$ and $\vect{t}\subset\vect{d}$, and if $\mathsf{O}=\sum_k \left( {\vect{y}}_k - {\vect{t}}_k \right)^2$, then, by using \eref{eq:PrototypeTimesPsi} for an arbitrary function $\vect{v}$,
\begin{equation}
  \begin{array}{lcl}
	\displaystyle \frac{\partial \mathsf{O}}{\partial \vect{p}_n} &=& \displaystyle 2 \sum_k \left( {\vect{y}}_k - {\vect{t}}_k \right) \cdot \frac{\partial {\vect{y}}_k}{\partial \vect{p}_n} = 2 \left( \vect{s} - \vect{t} \right) \cdot \frac{\partial \vect{s}}{\partial \vect{p}_n} =\\
	\\
	&=& \displaystyle 2 \left( \vect{s} - \vect{t} \right) \cdot \frac{\partial \vect{s}}{\partial \vect{p}_n} - \frac{\partial \matr{A}}{\partial \vect{p}_n} \vect{s} \cdot \vect{v} - \matr{A}\frac{\partial \vect{s}}{\partial \vect{p}_n} \cdot \vect{v} - \frac{\partial \vect{b}}{\partial \vect{p}_n} \cdot \vect{v} =\\
	\\
	&=& - \displaystyle \frac{\partial \matr{A}}{\partial \vect{p}_n} \vect{s} \vect{v} - \frac{\partial \vect{b}}{\partial \vect{p}_n} \vect{v},
  \end{array}
\end{equation}
if $\vect{v}$ is a solution of
\begin{equation}\label{eq:AdjointState2}
  \matr{A}^t\vect{v} = 2 \left( \vect{y} - \vect{t} \right).
\end{equation}
Therefore, it is clear the great computational advantage of obtaining the gradient of $\mathsf{O}$, with a single run of the FP to solve \eref{eq:AdjointState2}: the adjoint-state approach reduces both execution time and approximation errors with respect to a ``naive'' finite-differences strategy.  See \cite{ACKERER2014108} for an interesting application.

\section{Conclusions}

The conceptual framework developed in this paper is a generalization of those found in the literature on IP. It is very useful to introduce and formalize a series of topics that are of great relevance for geophysical applications.

Some relevant remarks are shortly summarised in this section.

The distinction between fixed and calibrated model parameters, together with the definitions of model prediction and model target, clearly highlights the role of experimental or monitoring data. Moreover, it also emphasizes the relevance of fixed parameters (e.g., the spacing of the discretization grid)  and of the data processing for inverse problems. In other words, there is no way to perform a satisfactory model calibration, without paying great attention to the available data, their accuracy, their physical consistency, their relevance at the model space- and time-scales.

Several difficulties are commonly encountered in inverse modelling and they are often claimed to be related to the ill-posedness of the IP. Actually, the real source of these problems should be carefully determined. The proposed framework is helpful to clarify and to point to the crucial critical aspects, which may affect the IP solution, as for example, the presence of several local minima of the objective function or the flatness of the objective function around the minimum. In these cases, changing the solution algorithms and testing different parameters (e.g., initialization of iterative minimization algorithms) of the applied algorithms might reduce the difficulties. However, researchers and professionals who are not experienced or well educated in inverse modelling, might incur in errors, misjudgement or oversight, if they do not fully control the solution methods, some of which could, for instance, span a small subspace of the whole parameter space.

A clear and rigorous definition of the IP is fundamental in order to properly analyse the posedness of the IP. Discussion in section \ref{sec:uniqueness} shows that a preliminary examination of identifiability might give very useful information about the IP, in particular with respect to uniqueness of the solution. Unfortunately, this is ignored in most applications of inverse modelling in geophysics. Moreover, uniqueness and stability of the IP are not easy to be assessed and, however, well-posedness would not be enough to obtain physically reliable values in practical applications. In fact, well-conditioning, which could be considered synonymous of robustness, is fundamental to guarantee that the errors introduced by data measurements and by the model approximations do not prevent from finding a reliable estimate of the model parameters.

The latter remark implicates a careful examination of the effects that uncertainties on the data and on the model outcome have on the IP solution. For the sake of brevity, the issue of resolution has not been considered extensively in this paper, but is briefly recalled here. Roughly speaking, if it is accepted that the discrepancy between $\vect{y}$ and $\vect{t}$ cannot be reduced at will, due to the great number of sources of errors, many of which were partly listed in section \ref{sec:StabilityAndConditioning}, then it should also be accepted that the values of fitting parameters could belong to a -- hopefully small -- region around the optimal array. This resolution issue is obviously linked, but is not equivalent, to stability and well-conditioning, and therefore needs a proper and dedicated analysis for practical applications.

The conceptual framework proposed in this paper is so complete as to permit to cast very different approaches in a unique framework and thus to facilitate the definition and the assessment of the intrinsic properties of the IPs, the effects of the methods of solution and to develop tools useful to perform further analysis on the results of inverse modelling.

%
%

\bibliographystyle{amsplain}


\providecommand{\bysame}{\leavevmode\hbox to3em{\hrulefill}\thinspace}
\providecommand{\MR}{\relax\ifhmode\unskip\space\fi MR }
\providecommand{\MRhref}[2]{%
  \href{http://www.ams.org/mathscinet-getitem?mr=#1}{#2}
}
\providecommand{\href}[2]{#2}

%
%

\end{document}